\documentclass{article} 
\usepackage{iclr2023_conference_tinypaper,times}

\usepackage{amsmath,amsfonts,bm}

\def\eqref#1{equation~\ref{#1}}

\def\1{\bm{1}}

\DeclareMathAlphabet{\mathsfit}{\encodingdefault}{\sfdefault}{m}{sl}
\SetMathAlphabet{\mathsfit}{bold}{\encodingdefault}{\sfdefault}{bx}{n}

\usepackage{hyperref}
\usepackage{url}
\usepackage{graphicx}
\usepackage{wrapfig}
\usepackage{xcolor}
\usepackage{caption}

\title{Analog In-Memory Computing with Uncertainty Quantification for Efficient Edge-based Medical Imaging Segmentation}

\author{Imane Hamzaoui \\
École nationale Supérieure d’Informatique \\
Algiers, Algeria\\
\texttt{ji\_hamzaoui@esi.dz} \\
\And
Hadjer Benmeziane\\
IBM Research Europe\\
8803 Rüschlikon, Switzerland \\
\texttt{hadjer.benmeziane@ibm.com} \\
\And
 Zayneb Cherif  \\
 Yorktown High school\\
 Yorktown Heights, 10598, USA \\
\texttt{zayneb.cherif@yorktown.org}
\And
Kaoutar El Maghraoui  \\
IBM T. J. Watson Research Center\\
Yorktown Heights, NY 10598, USA \\
\texttt{kelmaghr@us.ibm.com}
}

\iclrfinalcopy 
\begin{document}

\maketitle

\begin{abstract}
This work investigates the role of the emerging Analog In-memory computing (AIMC) paradigm in enabling Medical AI analysis and improving the certainty of these models at the edge. It contrasts AIMC's efficiency with traditional digital computing's limitations in power, speed, and scalability. Our comprehensive evaluation focuses on brain tumor analysis, spleen segmentation, and nuclei detection. The study highlights the superior robustness of isotropic architectures, which exhibit a minimal accuracy drop (0.04) in analog-aware training, compared to significant drops (up to 0.15) in pyramidal structures. Additionally, the paper emphasizes IMC's effective data pipelining, reducing latency and increasing throughput as well as the exploitation of inherent noise within AIMC, strategically harnessed to augment model certainty.


\end{abstract}

\section{Introduction}

Analog In-memory Computing (AIMC) marks a shift from traditional digital computing promising efficient and scalable processing for the rapidly growing medical data. Traditional digital systems, hindered by the Von Neumann bottleneck where data and instructions travel separately, struggle with large-scale data tasks, resulting in inherent inefficiencies. AIMC promises better efficiency and lower power use but faces challenges like susceptibility to noise, which can impact computation accuracy. In a recent study~\citep{bonnet-2023}, memristor-based Bayesian neural networks (BNNs) were investigated for heartbeats classification. In contrast, our work explores AIMC's application on a wider range of medical imaging tasks, analyzing if it can effectively address healthcare needs while managing noise issues. Our primary focus lies on the algorithmic aspects, given that AIMC accelerators are still in the early stages of development ~\citep{gallo-2023, wan-2022-rram,8894568-yin, 9731670-khwa-22}. The code is available via this link (\href{https://anonymous.4open.science/r/Analog_Med-B867}{https://anonymous.4open.science/r/Analog\_Med-B867}). 


\section{Evaluating Medical Deep learning on Analog IMC}

We present a comprehensive evaluation of Analog In-memory Computing (AIMC) for medical imaging, utilizing three benchmark datasets: Brain Tumor Segmentation, Spleen Segmentation, and Nuclei Detection. The study incorporates advanced architectures like U-Net~\citep{unet}, U-Net++~\citep{unetplusplus}, and Swin Transformer~\citep{swin}, trained via AIHWKIT~\citep{aihwkit-DBLP:journals/corr/abs-2104-02184}.  The benchmarks and training methodologies are described in the appendix \ref{appendix:datasets} and appendix\ref{appendix:training}. This analysis aims to assess AIMC's effectiveness in critical medical imaging tasks, highlighting its potential and capabilities in this evolving field. 



\begin{figure}
    \centering
    \vspace{-0.5cm}
    \includegraphics[width=\textwidth]{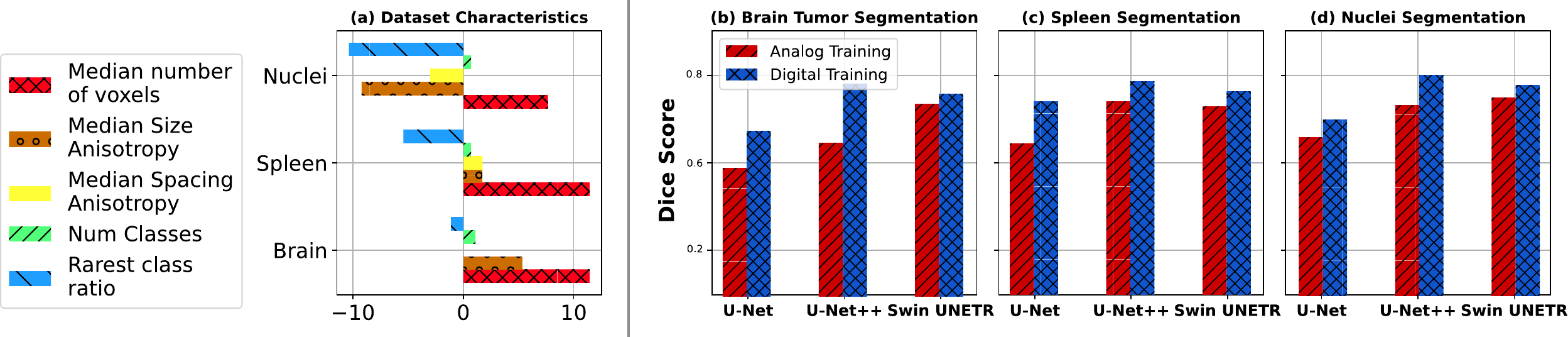}
    \caption{(a) Dissimilarity and targeted task variance. b-c-d) Noise-induced dice score drop in different medical datasets.}
    \label{fig:overall-analog}
\end{figure}

{{\textbf{Noise-resiliency in medical imaging analysis models:}}}\,We depict the diversity of the benchmarks we evaluate in Figure~\ref{fig:overall-analog}(a), highlighting various characteristics such as the size, the number of classes, etc., inspired by~\citep{isensee-2020-nnunet}. 

Our study on noise resilience in medical imaging models highlights two key insights. First, the pyramidal structure of U-Net models leads to increased noise vulnerability, as evidenced by dice score reductions of 0.15 and 0.22 for U-Net and U-Net++ respectively in brain segmentation tasks (see Figure~\ref{fig:overall-analog}). This vulnerability is attributed to their alternating down-sampling and up-sampling design which can amplify noise variations. In contrast, Swin-like transformer architectures exhibit remarkable noise resilience, with a negligible 0.04 performance drop as shown in Figure~\ref{fig:overall-analog}. Their isotropic design, which treats image patches consistently and lacks hierarchical convolutional operations, contributes to their enhanced stability against noise disturbances.


\begin{table}[h]
    \centering
    \begin{tabular}{p{5.5cm}|p{2cm}|p{2.5cm}|p{2.5cm}}
    \hline
         Model & Avg tile utilization (\%) & Avg Reuse factor & \# Parameters (M) \\ \hline\hline
         U-Net~\citep{unet} & 7.42   & 6574.7 & 7.76  \\ \hline
         UNet++~\citep{unetplusplus} & 12.53 & 8721.0 & 19.6  \\ \hline
         Swin UNET~\citep{swin} & 43.2  & 11540.6 &  62.19  \\ \hline
    \end{tabular}
    \caption{Performance metrics of state-of-the-art medical imaging models.}
    \label{tab:hw_perf}
\end{table}

{\textbf{Model Inference on MRIs \& CT images:}}\,
In medical imaging, such as MRI and CT scans, Analog In-memory Computing (AIMC) significantly enhances data processing efficiency. Unlike traditional methods, AIMC's pipelining ability allows for rapid, parallel processing of sequential image slices, crucial for three-dimensional anatomical analysis. This approach not only improves throughput in urgent medical scenarios but also optimizes energy usage and minimizes latency between slices, making it particularly effective for volumetric data in tumor segmentation.

\begin{figure}[h]
    \begin{minipage}[b]{0.15\textwidth}
        \captionsetup{justification=raggedright, singlelinecheck=false}
        \caption{Uncertainty analysis on sample brain tumor segmentation images for UNet++.}
        \label{fig:uncertain_examples}
    \end{minipage}
    \hfill 
    \begin{minipage}[b]{0.8\textwidth}
        \centering
        \includegraphics[width=\textwidth]{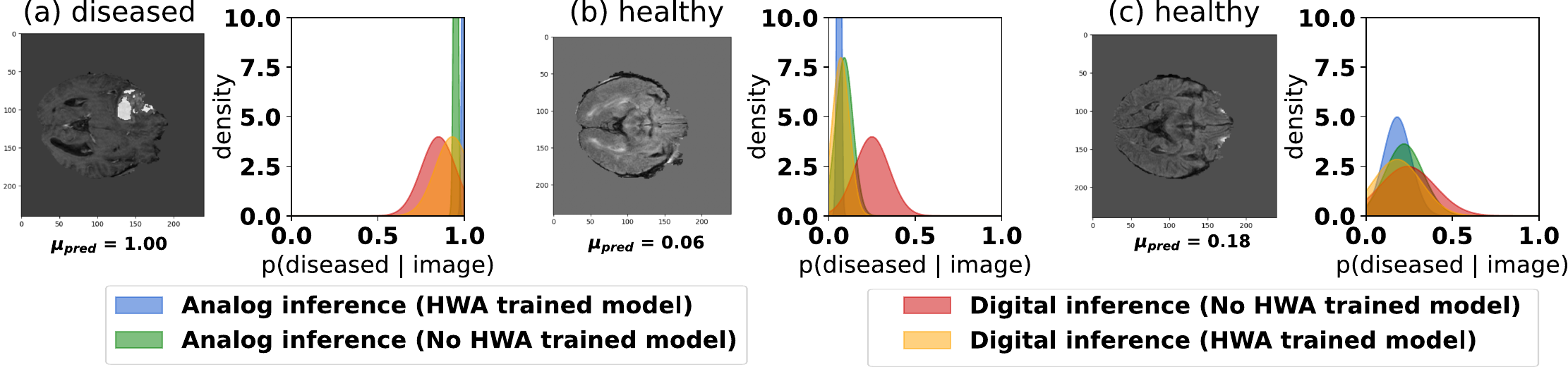}
    \end{minipage}
\end{figure}
{\textbf{Certainty enhanced through noise:}}\,
While noise in computational models is often seen as detrimental, in hardware-aware training (HWA training), it inadvertently leads to more resilient models. These models, trained under controlled noise conditions, show improved noise tolerance and prediction certainty, as visualized in Figure~\ref{fig:uncertain_examples}. In critical healthcare applications, the certainty of model predictions is vital, as it minimizes the risk of misdiagnosis and enhances decision-making in treatment planning. This is especially evident when comparing digital and analog-trained models, such as the U-Net++ architectures. 

    

%

\section{Conclusion}
In-memory computing (IMC) holds promise in refining medical imaging, with transformer structures surpassing their pyramidal counterparts in resilience to noise. Rather than being detrimental, strategic noise injection fortifies model precision, an essential aspect in healthcare. This approach mitigates overfitting and boosts confidence in diagnostics. Future efforts will focus on advancing transformer-based analog-aware architectures through the application of neural architecture search, catering to a wide array of medical imaging tasks.
\section*{Acknowledgment}
 This research work was supported by Abu Dhabi National Oil Company (ADNOC), Emirates NBD, Sharjah Electricity Water \& Gas Authority (SEWA), Technology Innovation Institute (TII) and GSK as the sponsors of the 4th Forum for Women in Research (QUWA): Sustaining Women’s Empowerment in Research \& Innovation at University of Sharjah. 

\bibliography{iclr2023_conference_tinypaper}
\bibliographystyle{iclr2023_conference_tinypaper}

\newpage
\appendix
\section{Appendix}
\begin{figure}
    \centering
    \includegraphics[width=\textwidth ]{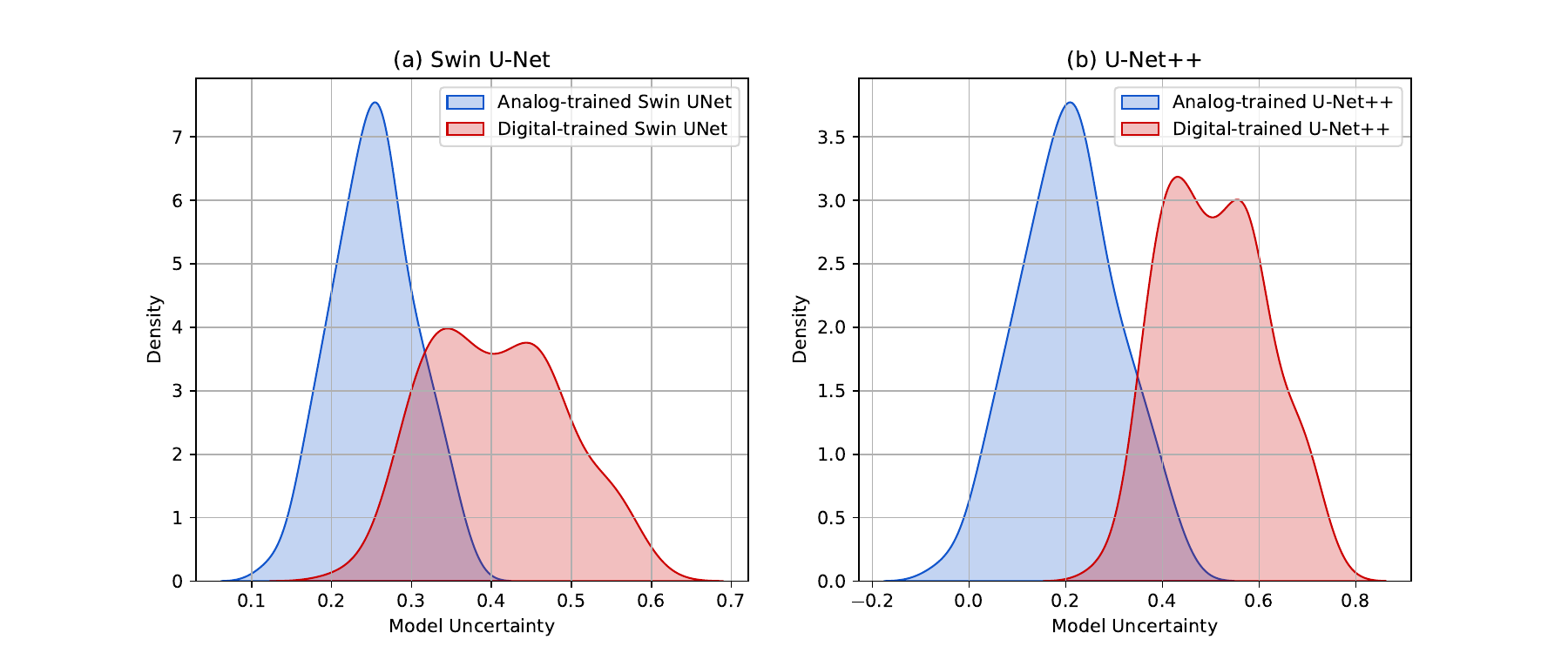}
    \caption{Model uncertainty density analysis across different brain tumor detection inputs.}
    \label{fig:uncertainty_model}
\end{figure}
\subsection{Datasets}\label{appendix:datasets} 
The Brain Tumor Segmentation dataset from The Cancer Imaging Archive~\citep{decathlon_dataset} features MRI scans and segmentation masks for 110 lower-grade glioma patients, enriched with FLAIR sequences and genomic cluster data. The Spleen Segmentation dataset, also from the Medical Segmentation Decathlon~\citep{decathlon_dataset}, offers CT scans for detailed spleen segmentation, focusing on organ delineation. Lastly, the Nuclei detection dataset~\citep{caicedo-2019} provides a broad collection of segmented nuclei images, showcasing diversity in cell types and imaging modalities, including brightfield and fluorescence techniques.

\subsection{Models \& training} \label{appendix:training}
U-Net, effective in biomedical segmentation, offers a balanced 'U' shaped structure for localization. U-Net++ enhances U-Net by adding nested and skip pathways for improved detail capture. Swin Transformer, unlike these, utilizes shifted windows to handle image patches, focusing on long-range interdependencies. Transitioning from initial training, the models were adapted to hardware-aware training through noise injection, facilitated by AIHWKIT~\citep{aihwkit-DBLP:journals/corr/abs-2104-02184}. This approach, supported by MONAI~\citep{cardoso2022monai} frameworks and other open-source tools, optimized our models for in-memory computing applications. The simulation of the analog aware training's parameters and characteristics is available in the provided code and defined in the function $create\_rpu\_config()$.

\subsection{Extended Certainty analysis}

Figure~\ref{fig:uncertainty_model} provides an overall certainty analysis of U-Net++  and Swin U-Net when trained with hardware training versus digital training. While the main paper presents a compelling and illustrative example, we extend the density quantification to the whole test set. The uncertainty is computed using Monte Carlo Sampling. This involves using multiple passes of the input data and observing the variability in the outputs. The lower the variability, the higher the certainty of the model's predictions. 

Results suggest that the analog-trained model exhibits a lower uncertainty density compared to the digital-trained model. This implies that the analog-trained U-Net++ is be more reliable and consistent in its predictions, making it a more suitable choice for medical applications. This is mainly due to the hardware-aware training that forces the model to adapt to the inherent variability and constraints of the analog computing environment, thereby enhancing its ability to handle uncertain scenarios with greater precision. 

\end{document}